\documentclass[twocolumn, prl, superscriptaddress,
 amsmath,amssymb,
]{revtex4-2}

\usepackage{hyperref}
\usepackage{xcolor}

\usepackage{graphicx}
\usepackage{dcolumn}
\usepackage{bm}

\usepackage{amssymb}

\begin{document}

\preprint{APS/123-QED}

\title{Emergence of Purely Elasto-Plastic Turbulence in Shear Flows} 

\author{Muhammad Abdullah}
\affiliation{%
 Department of Mechanical Engineering and Applied Mechanics, University of Pennsylvania, Philadelphia, PA 19104
}%

\author{Shravan Pradeep}
\affiliation{%
 Department of Mechanical Engineering and Applied Mechanics, University of Pennsylvania, Philadelphia, PA 19104
}
\affiliation{%
 Department of Earth and Environmental Science, University of Pennsylvania, Philadelphia, PA 19104
}%

\author{Doug J. Jerolmack}
\affiliation{%
 Department of Mechanical Engineering and Applied Mechanics, University of Pennsylvania, Philadelphia, PA 19104
}
\affiliation{%
 Department of Earth and Environmental Science, University of Pennsylvania, Philadelphia, PA 19104
}%

\author{Becca Thomases}
\affiliation{%
 Department of Mathematics, Smith College, Northampton, MA 01063}
 
\author{Paulo E. Arratia}%
\altaffiliation{Corresponding author}
\email{parratia@seas.upenn.edu}
\affiliation{%
 Department of Mechanical Engineering and Applied Mechanics, University of Pennsylvania, Philadelphia, PA 19104
}%

\date{\today}

\begin{abstract}
We observe the emergence of a distinct, elasticity-driven flow state in a yield-stress fluid in the absence of inertia. Numerical simulations show that this elasto-plastic turbulent state is characterized by a broad spectrum of fluctuations in velocity and stress. Results show a non-monotonic relationship between the volume fraction of the unyielded flow and plasticity. Surprisingly, we find that above a critical value of plasticity, the system can fluidize. Our results reveal the complex interplay between elasticity and plasticity in simple shear flows, indicating that plasticity can enhance rather than hinder momentum transport. 
\end{abstract}

\maketitle


Yield-stress fluids behave like a solid, but flow like a liquid above a critical stress level \cite{bingham1917investigation,Bonn_Science,larson99}. They are widely found in nature and industrial processes \cite{bingham1917investigation,Bonn_Science,Ewoldt_ARFM,bonn2017yield,Balmforth_ARFM, COUSSOT_JNNFM_2014,SimonRogers_PRL2021}; examples include particle suspensions and emulsions, gels and foams, clays and muds, and cosmetics and food products \cite{larson99,bonn2017yield,Balmforth_ARFM,pradeep2024origins}. Yield-stress fluids often exhibit nonlinear rheological behavior, with complex spatio-temporal dynamics governed by the simultaneous influence of viscosity, elasticity, and plasticity \cite{SARAMITO20071,SimonRogers_PRL2021,Coussot_PRL2019}. Below yielding, they can behave elastically, deforming and storing energy; above yielding, their flow is that of a viscoelastic fluid. This class of yield-stress materials are the so-called elasto-viscoplastic (EVP) fluids. The material and rheological behavior of EVP fluids have been the subject of much work \cite{coussot2007rheophysics,malkin2017modern,bonn2017yield, frigaard2019simple, ovarlez2013existence, Coussot_PRL2019, Barrat_RevModPhys2018}, as has the development of constitutive models describing their elastic behavior \cite{SARAMITO20071,saramito2016complex}, dating back to even the early 1900s \cite{schwedoff1900rigidite,oldroyd1947rational}. Despite these efforts, the combined nonlinear effects of plasticity (yield-stress) and elasticity on fluid flow remain poorly understood.  


The effects of elastic stresses on flows of yield-stress fluids have only recently begun to be elucidated. At high Reynolds numbers (Re), in the weakly elastic regime, numerical simulations using the Saramito EVP constitutive model \cite{SARAMITO20071} that includes elastic effects show that plasticity strongly impacts turbulent channel flow, driving laminarization and enhanced intermittency \cite{Rosti_Izbassarov_Tammisola_Hormozi_Brandt_2018}. The dynamics also feature a highly unsteady yielding/unyielding process along the centerline. Modal analyses further detail the organization of coherent structures in turbulent EVP channel flows \citep{LeClainche2020EVPStructures}; similar trends are also observed in two-dimensional turbulence \cite{Rosti_Nature}. As elastic effects become more dominant, direct numerical simulations of turbulent channel flows show that EVP fluids can exhibit a significant reduction in drag relative to Newtonian and weakly elastic cases \cite{Izbassarov_Rosti_Brandt_Tammisola_2021}. 


In the low Re regime, simulations and experiments have reported symmetry breaking, flow fluctuations, and negative wakes in EVP fluid flows around obstacles \cite{cheddadi2011understanding,fraggedakis2016yielding}. Recently, it has been argued that plasticity suppresses the onset of elastic turbulence in the canonical four-roll mill extensional flow geometry \cite{Dzanic}. In the same low-inertia limit -- hereafter dubbed the \textit{elasto-plastic} regime -- simulations show that the volume fraction (i.e., the proportion of unyielded fluid) in an EVP shear (i.e., Kolmogorov) flow monotonically increases with plasticity \cite{Jamming2025}. 
Surprisingly, elastic stresses have been observed to increase the size of unyielded flow regions and, consequently, the pressure drop in EVP flows across porous media \cite{chaparian_yield-stress_2020,Hormozi_JNNFM_Porous_2018,parvar_general_2024}. Experiments, on the other hand, find that elasticity reduce unyielded regions and induce time-dependent behavior \cite{abdelgawad_interplay_2024}; in particular, normal stresses often associated with elastic behavior, are found to grow chaotically after yielding \cite{venerus2022evidence}. These results highlight the intricate and still poorly understood interplay between elasticity and plasticity in EVP fluid flows, despite their critical role in natural \cite{JerolmackandDaniels} and industrial \cite{Ewoldt_ARFM} processes.

In this manuscript, we numerically investigate the effects of plasticity on purely elastic turbulence in shear flows. Here, plasticity refers to the fluid nonlinear yielding behavior.  As plasticity is increased beyond a critical value, we find the emergence of an elasto-plastic turbulent state characterized by a broad range of temporal fluctuations in the kinetic and strain energies (see Fig. 2). The transition to this new randomized flow state exhibits periodic behavior, marked by a collapse of the broad elastic turbulent spectra into limit cycles. 
As plasticity becomes increasingly dominant relative to elasticity, the mean unyielded volume fraction decreases, while velocity fluctuations grow stronger. Our results provide evidence that plasticity (yield-stress), when coupled with elastic effects, may promote rather than arrest momentum and energy transport.


\begin{figure}[t]
    \centering
    \includegraphics[width = 0.5\textwidth]{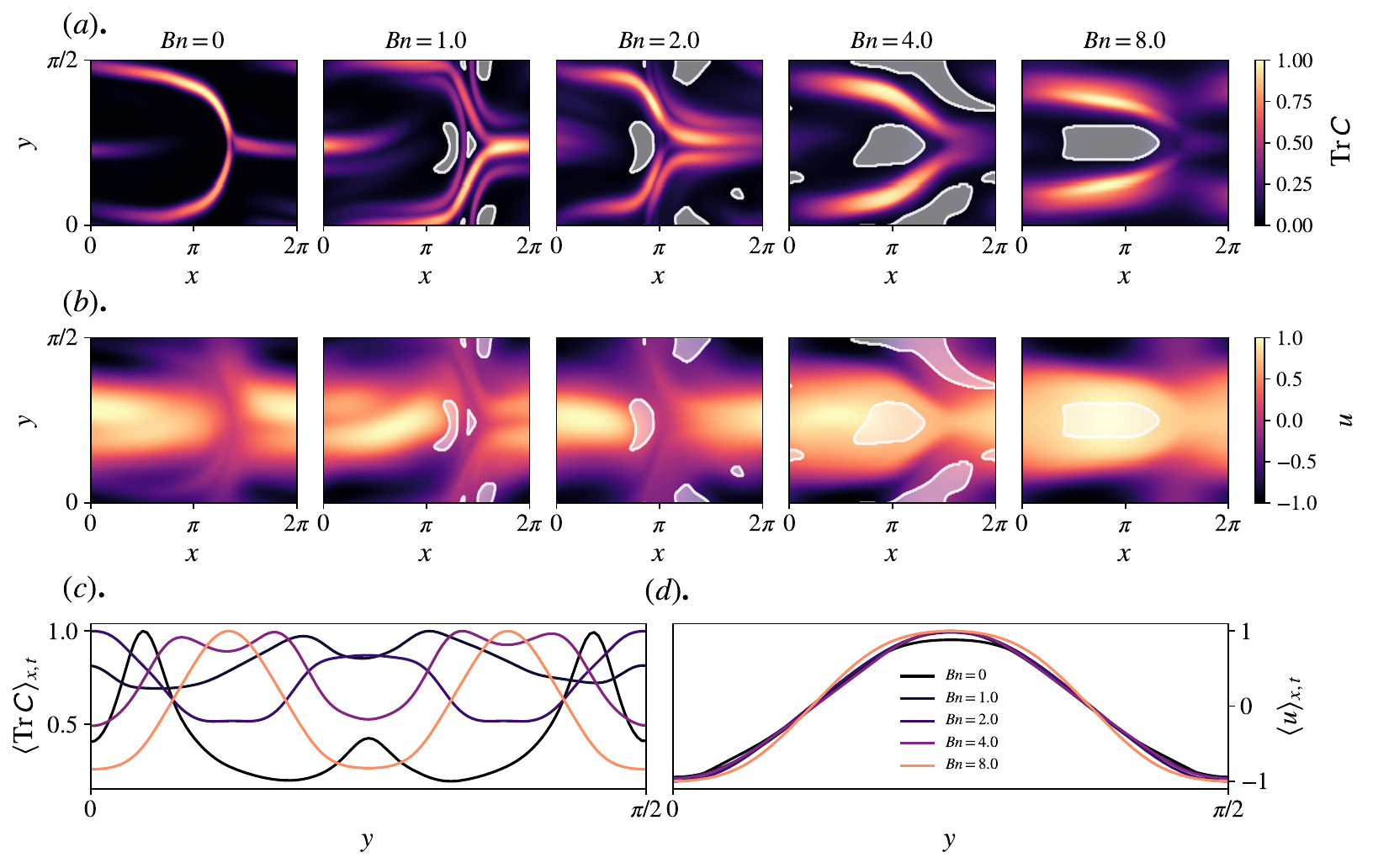}
    \caption{For $Bn\in\left\{0, 1, 2, 4, 8\right\}$, representative instantaneous profiles of (a): the trace of the conformation tensor $\mathrm{Tr}\,\mathsf{C}$ and (b): the streamwise velocity $u$, each normalized by the absolute maximum; note that $Bn = 0$ denotes the purely viscoelastic case. In each panel, the white regions represent unyielded fluid. Spatio-temporal averages (along $x$ and time, denoted by $\left\langle\cdot\right\rangle_{x,t}$) are plotted versus the wall-normal coordinate $y$ for  $(c)$, $\mathrm{Tr}\,\mathsf{C}$ and $(d)$, $u$, again normalized by the absolute maximum. See Fig. 1 in SM for unnormalized behavior.}
    \label{fig:fig1}
\end{figure}

Our system consists of a doubly periodic Kolmogorov flow of an elastoviscoplastic (EVP) fluid. This is a benchmark flow because of its interpretable dynamics and significance as a minimal unit for the more complex channel flow. On the numerical domain $\left(x, y\right)\in\left[0, 2\pi\right]\times\left[0, 2\pi/n\right]$, with $n = 4$, we adopt the Saramito model \cite{SARAMITO20071} to describe the flow of an EVP fluid in the Stokes' regime
\begin{align}
\label{eqn:conf_eqs}
    \overset{\nabla}{\mathsf{C}} & = -\mathcal{M}\left(Wi, Bn, \mathsf{C}\right)\left(\mathsf{C} - \mathsf{I}\right) + \nu \nabla^2\mathsf{C}, \\
    -\nabla p + \nabla^2 \boldsymbol{u} & = -\left(\xi/\lambda\right)\left(\nabla\cdot\mathsf{C}\right) + \boldsymbol{f}, \\
    \nabla\cdot\boldsymbol{u} & = 0.
\end{align}
Here, $\boldsymbol{u} = \left(u\,\,\,v\right)$ is the Eulerian velocity,  $\xi = \eta_p/\eta_s$, where $\eta_p$ and $\eta_s$ are the polymer and solvent viscosities, and $\mathsf{C}$ is the conformation tensor. The latter relates in turn to the non-Newtonian stress $\mathsf{T}$ through $\mathsf{T} = (\eta_p/ \lambda)\left(\mathsf{C}-\mathsf{I}\right)$, where $\lambda$ is the fluid relaxation time.
 Polymer diffusion 
  $\nu$ is added for numerical stability \cite{Becca_PRL, Alves_ARFM_Viscoelastic, Gupta_Vincenzi_2019,Dzanic_ET_PRE, morozov_PRL}. 
The dimensionless numbers that govern this flow system are the Weissenberg number, $Wi = \lambda U/L$, and the Bingham number, $Bn = T_y L/\eta_p U$, where $T_y$ is the yield-stress and $U$ and $L$ are the velocity and length scales. As defined, $Bn$ describes the relative importance of yield-stress (plasticity) to elastic stresses. 
Finally, a sinusoidal forcing $\boldsymbol{f}$ prescribes the periodic Kolmogorov flow. Please see numerical methodology section in SM for more details.

In Equation (\ref{eqn:conf_eqs}), $\mathcal{M}$ is defined
as $\mathcal{M}\left(Wi, Bn, \mathsf{C}\right) = \max\left(0, 1 - Wi Bn/C_d\right)/\lambda$, where $C_d = \sqrt{\left(\mathsf{C}_d\colon\mathsf{C}_d\right)/2}$, the magnitude of the deviatoric contribution $\mathsf{C}_d$ to $\mathsf{C}$. For the Saramito model, $0\leq\mathcal{M}\leq 1$ reflects the yielding response of the fluid; if $\mathcal{M} = 0$, the material is unyielded (a Kelvin-Voigt viscoelastic solid). For $\mathcal{M} > 0$, the fluid is yielded, and we recover features of the standard Oldroyd-B model (with $\mathcal{M} \to 1$ obtaining it exactly). Since the goal is to investigate the effects of plasticity on chaotic viscoelastic flows, we set $Wi = 28$ for all simulations; this value of $Wi$ has been shown to elicit elastic turbulence in the purely viscoelastic case (i.e., $Bn=0$) \cite{Becca_kolm}.  


\begin{figure*}[t]
    \centering
    \includegraphics[width = \textwidth]{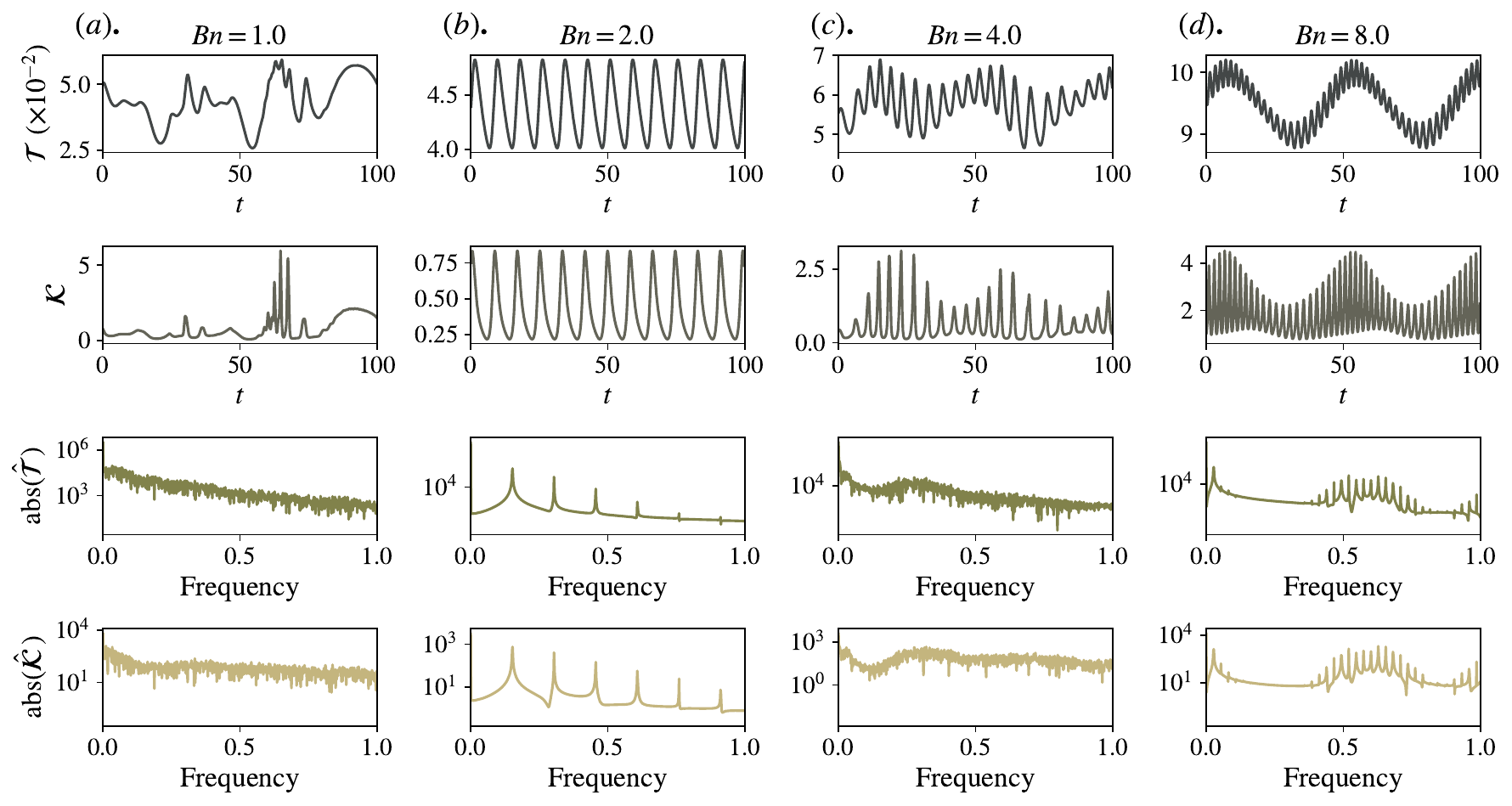}
    \caption{From the top row to bottom, plots of the domain-averaged strain and kinetic energies, the volume fraction $\phi$, and the strain energy spectra for $(a)$, $Bn = 1$, $(b)$, $Bn = 2$, $(c)$, $Bn = 4$ and $(d)$, $Bn = 8$. }
    \label{fig:fig2}
\end{figure*}

We begin by analyzing the structures produced by EVP fluids in shear flows. Figures \ref{fig:fig1}$(a)$ and $(b)$ show snapshots of, respectively, the trace of the conformation tensor $\mathrm{Tr}\,\mathsf{C}$ and the streamwise velocity $u$ as a function of $Bn$, ranging from the purely elastic case ($Bn=0$) to a case where the yield-stress dominates the elastic stresses ($Bn=8$). The quantity $\mathrm{Tr}\,\mathsf{C}$ encodes information on the distribution of non-Newtonian stresses but also represents a form of strain energy. A hallmark of viscoelastic shear flows is the emergence of traveling-wave coherent structures in the shape of  \textit{narwhals} \cite{morozov_PRL,Becca_kolm} or \textit{arrowheads} \cite{arrowheads_dubief}. These structures have been argued to contribute to organizing and mobilizing the dynamics in viscoelastic shear flows \cite{Becca_kolm, morozov_PRL, Lewy_Kerswell_2025, Morozov_PNAS_2024}. Figure \ref{fig:fig1}(a) shows that such coherent structures persist despite the increase in $Bn$, albeit with a morphology strongly influenced by plasticity. More specifically, for larger values of $Bn$, the stresses seem to organize themselves around a central unyielded plug that tends to taper and thicken the structure ``body" while simultaneously suppressing its centerline. 
This is also apparent in the distributions of the spatio-temporally averaged profiles for $\mathrm{Tr}\,\mathsf{C}$ in Fig. 1$(c)$, which shows how plasticity significantly affect the flow coherent structures (non-Newtonian stresses). See movies in SM.

 
The streamwise velocity fields in Fig.~\ref{fig:fig1}$(b)$ show the gradual appearance of unyielded regions in the flow as $Bn$ increases. Similarly to $\mathrm{Tr}\,\mathsf{C}$, as $Bn$ is increased further, these unyielded regions organize themselves into a larger (coherent) structure at the center of the streamwise velocity field. 
This behavior has been observed in experiments, although in a different geometry \cite{abdelgawad_interplay_2024}. Despite this flow reorganization, the spatio-temporally averaged streamwise velocity $\bar{u}\equiv \left\langle u\right\rangle_{x,t}$ remains nearly unchanged (Fig.~\ref{fig:fig1}$(d)$) as $Bn$ varies, maintaining an approximately sinusoidal form consistent with previous work \cite{boffeta_2005_a, boffeta_2005_b, berti_boffetta,SARAMITO20071}. It is surprising, however, that minimal differences in $\bar{u}$ can be identified across $Bn$, despite significant variations in $\mathrm{Tr}\,\mathsf{C}$. Interestingly, the locations of the unyielded regions in Figures \ref{fig:fig1}$(a)$ and $(b)$ highlight a preferential arrangement around the coherent structure's body (a traveling plug-like structure, see movies in SM). It is likely that the shape of this structure is a consequence of the changes in the stress coherent structure discussed earlier. Overall, these results show how plasticity can reorganize the structure of the instantaneous velocity and stress fields. 


We now investigate the flow dynamical behavior by examining the temporal signals of the domain-averaged strain ($\mathcal{T}\left(t\right) = (1/L_xL_y)\iint \mathrm{Tr}\,\mathsf{C}\,\mathrm{d}x\,\mathrm{d}y$) and kinetic energies ($\mathcal{K}\left(t\right) = (1/L_xL_y)\iint \left
\lVert\boldsymbol{u}\right\rVert^2\,\mathrm{d}x\,\mathrm{d}y$), as well as their respective spectra (Fig. \ref{fig:fig2}). 
%
Note that our control case, $Bn = 0$, is an elastic turbulent state characterized by broad temporal fluctuations \cite{Becca_kolm}. As shown in Fig. \ref{fig:fig2}(a), we find no significant changes in dynamics up to $Bn=1$ (see also Fig. 2 in SM); in this regime, elasticity dominates the dynamics. 

A flow transition appears as the relative importance of plasticity (yield-stress) overcomes elasticity (normal stresses). In this regime, $1<Bn< 4$, the wide spectrum of frequencies observed at $Bn<1$ collapses into a limit cycle, as shown in Figure \ref{fig:fig2}$(b)$. The data (see also the movies in the SM) reveal that the stress-coherent structure typical of ET no longer propagates as a traveling wave; instead, it undergoes a slow, spatially localized undulation within the domain. We note that similar behavior is found in the ET flow system as $Wi$ is decreased (that is, $Wi<28$) \cite{Becca_kolm}. This suggests that viscous dissipation from plasticity may be responsible for the development of the observed fundamental elasto-plastic frequency in the spectra. 


Increasing plasticity further (i.e., $Bn \approx 4$) leads to a surprising return to randomized dynamics (Fig. \ref{fig:fig2}$(c)$). Unlike the elastic turbulent state, however, there seems to be a dominant low frequency that emerges. We conjecture that the high-frequency oscillations are associated with the release of elastic energy, whereas the low-frequency oscillations (seen in Fig. 2$(b)$) are set by plastic dissipation. Of course, the combined (nonlinear) action of those effects leads to the observed spectra. Thus, we call this an elasto-plastic turbulent (EPT) state. 


At relatively high Bingham numbers, $Bn \geq 8$ (Fig. \ref{fig:fig2}$(d)$), we find that the elasto-plastic turbulent signals are modulated by a slow (plastic) temporal oscillation. The flow becomes organized by the stress coherent structures, and features return to their traveling-wave-like state; these structures translate (mostly) in the streamwise direction, although with rapid unsteady changes in direction (see movies in SM). 

\begin{figure}[t]
        \includegraphics[width=0.4\textwidth]{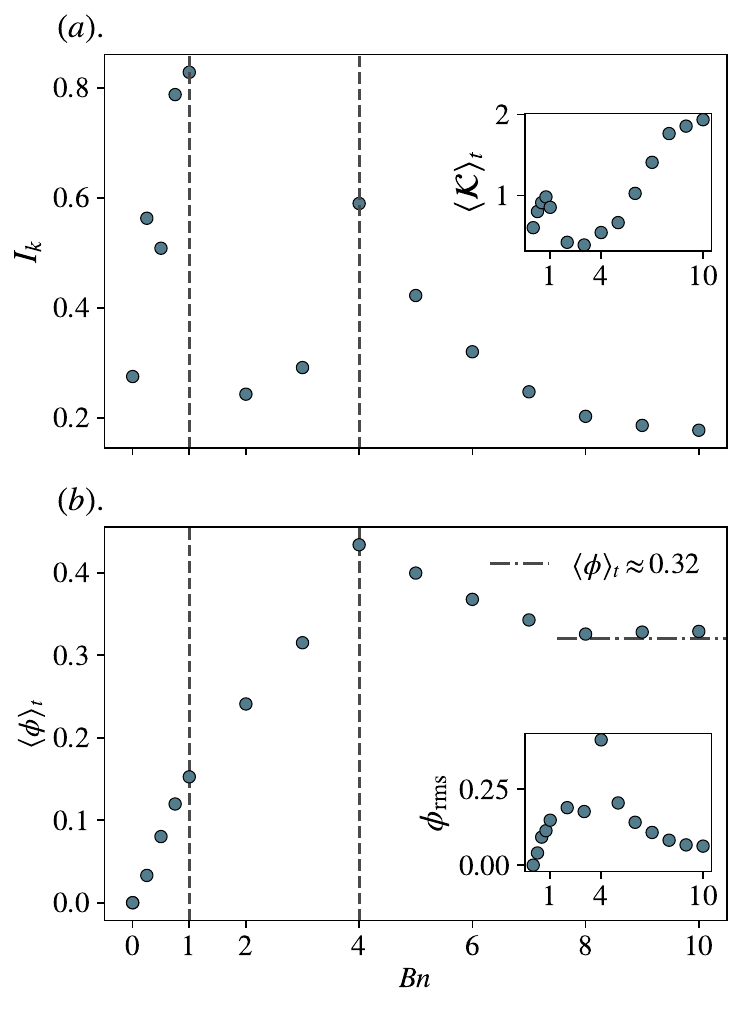}
        \caption{(a) $I_k$, the magnitude of the velocity fluctuations normalized by the mean energy as a function of $Bn$, and the temporally-averaged kinetic energy (inset). $(b)$ The time-averaged volume fraction $\left\langle \phi\left(t\right)\right\rangle_t$ as a function of increasing plasticity, captured through non-dimensional Bingham number ($Bn$), show the emergence of a critical plasticity around critical Bingham number, $Bn_c \simeq 4$. The inset shows values of the root-mean-square of $\phi$ fluctuations ($\phi_{\mathrm{rms}}$) as a function of increasing $Bn$. }
        \label{fig:fig3}
\end{figure}

\begin{figure}[t]
    \centering
    \includegraphics[width = 0.4\textwidth]{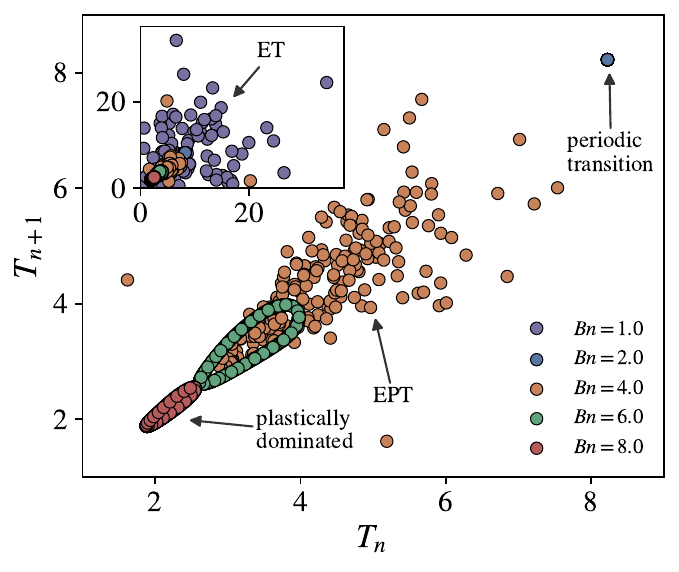}
    \caption{The period return map $\left(T_n, T_{n+1}\right)$ in $\mathcal{T}$ for $Bn=2,4,6,8$, and with $Bn=1$ (inset). The map show the dynamical flow transition as a function of $Bn$.}
    \label{fig:fig4}
\end{figure}

We further characterize these flow states by examining the velocity fluctuations of the base ET flow as a function of plasticity ($Bn$). Figure \ref{fig:fig3}$(a)$ shows the quantity $I_k\equiv\mathcal{K'}/\mathcal{K}_{\mathrm{mean}}$, the energy of the fluctuations $\mathcal{K}^{\prime}$ normalized by the flow field mean energy $\mathcal{K}_\mathrm{mean}$, as a function of $Bn$. Our results show that the addition of plasticity to the ET flow leads to an increase in velocity fluctuations ($I_k$) with a pronounced peak at $Bn=1$. In this viscoelastic-dominated regime ($Bn\leq1$), velocity fluctuations arise from deformation and reorientation of polymer molecules due to velocity gradients. Thus, the increase in $I_k$ is associated with the increase in local shear-rates due to the appearance of small but finite unyielded regions in the flow domain that grow monotonically up to $Bn=1$  (see Fig. \ref{fig:fig3}a). In fact, we observed a corresponding monotonic increase in flow speed, $\langle\mathcal{K}\rangle_t$, with $Bn$ (Fig. \ref{fig:fig3}a, inset). As plasticity begins to dominate the system ($1<Bn< 4$), the flow transitions into the periodic state characterized by a sudden decrease in $I_k$ and $\langle\mathcal{K}\rangle_t$, likely due to viscous dissipation and flow structural re-organization (Fig. \ref{fig:fig1}).  

Surprisingly, for $Bn>3,$ both $I_k$ and $\langle\mathcal{K}\rangle_t$ reverse trend and begin to grow again, marking the emergence of the EP turbulent flow state seen at $Bn\approx4$. The mechanism for the sudden increase in $I_k$ is unclear, but it is likely due to the interplay between the elastic relaxation process ($\lambda$) and the viscous dissipation time-scale ($\eta/T_y$); qualitative evidence of these two times scales at work can be seen in Fig. \ref{fig:fig2}$(c)$, particularly in the strain energy signal. Deep into the plastic-dominated regime, for $Bn>4$, we find a decrease in $I_k$ due to relatively large viscous dissipation. Overall, this data shows that the elasto-plastic turbulent state is characterized by 
relatively large velocity fluctuations, similar in scale to those in the elastic turbulent regime, due to the nonlinear interplay between elastic stresses and plasticity.


Next, we investigate the growth of (un)yielded flow regions as a function of $Bn$. It would be reasonable to expect that unyielded regions would grow as plasticity (yield-stress) increases. However, a more nuanced picture emerges. Figure \ref{fig:fig3}$(b)$ shows the time-averaged unyielded area fraction $\left\langle \phi\left(t\right)\right\rangle_t$ as a function of $Bn$. Here, $\phi\left(t\right) = N_{\mathcal{M} = 0}/\left(N_xN_y\right)$ \cite{Jamming2025}, where $N_{\mathcal{M} = 0}$ denotes the cardinality of the set of grid points over which $\mathcal{M}$ vanishes and so, the fluid is unyielded. As expected, $\phi$ increases monotonically with plasticity up to $Bn = 4$; a similar trend was previously reported in numerical simulations of EVP flows in porous media \cite{Jamming2025}. Surprisingly, a counterintuitive decrease in $\left\langle \phi\left(t\right)\right\rangle_t$ is observed for $Bn > 4$. That is, as plasticity increases, the system seems to fluidize again. The data ultimately plateaus at $\left\langle \phi\left(t\right)\right\rangle_t\approx 0.32$ for $8 \le Bn \le 10$.  Similar behavior is found by computing the root-mean-square of the volume fraction fluctuations  $\phi_\mathrm{rms} = \langle \left(\phi - \langle \phi\rangle_t\right)^2\rangle_t$ as a function of $Bn$ (Fig. \ref{fig:fig3}$(b)$, inset). The non-monotonic relationship between $\phi$ and $Bn$ is reminiscent of critical behavior in condensed matter physics, with unyielding volume fraction $\phi$ as the order parameter. Data shown in Fig. \ref{fig:fig3} suggests the existence of a critical point at $Bn_c \approx 4$; for $Bn<4$, the size and fluctuations of unyielded regions increase monotonically with plasticity; while a decreasing trend is found for $Bn>4$. The divergence around the critical is clear as $Bn_c$ is approached from high plasticity, as shown in root mean fluctuations of $\phi$ (Fig. \ref{fig:fig3}(b), inset). 
It is unclear and perhaps unlikely that the portion of the flow that remains unyielded will remain constant as $Bn$ is further increased, since plasticity is expected to eventually percolate the entire flow system for large enough values of $Bn$.  Nevertheless, the nonlinear dependence of $\phi$ with $Bn$ reveals that, in the presence of elasticity, plasticity can enhance, rather than hinder, momentum transport. 

Finally, Fig. \ref{fig:fig4} explores flow dynamical correlations by considering the return map of the period of the strain energy ($T_{n+1}$ versus $T_n$). As expected, these pairs display no discernible organization for the ET-dominated state ($Bn=1$, inset). 
During the transitional regime, ($Bn=2$), these pairs follow a convergent trajectory towards a unique period. In the elasto-plastic turbulent (EPT) state ($Bn= 4$), there is again much less coherence in the period return map, but we do see a tighter bound on the range of periods than in the ET regime. This bound continues to tighten and collapse onto thin ellipses for 
%
%
 higher ($Bn=6,8$), which suggests that the signal is now coupling the fast EP frequency with a slow (plastic) mode. This is confirmed by the spectrum  (see Fig. \ref{fig:fig2} $(d)$) where  
clusters appear around integer multiples of the slow frequency with sidebands spaced nearly uniformly by the fast frequency. 
Physically, this is indicative of the plastic state modulating the strength and rate of the underlying fast flow oscillation characteristic of elasto-plastic turbulence.


In summary, simulations show the emergence of a chaotic flow state in a simple shear flow of an elastic yield-stress fluid at low Reynolds number. This new elasto-plastic turbulent (EPT) state is characterized by a broad range of frequencies in kinetic and strain energies (Fig. \ref{fig:fig2}) and relatively large velocity fluctuations (Fig. \ref{fig:fig3}a). The transition to the EPT flow state seems to occur at $Bn\approx4$; the transition state shows flow re-organization and strong periodicity (Fig. \ref{fig:fig2}b and Fig. \ref{fig:fig4}) and is distinct from the randomized dynamics characteristic of elastic turbulence (Fig. \ref{fig:fig2}c). Surprisingly, we find that the unyielded flow volume fraction ($\phi$) decreases as plasticity is increased above the critical value of $Bn=4$ (where EPT flow state is found). The approach to $Bn_c$ from both sides, especially close to the critical point, is of interest but needs to be further investigated. Nonetheless, these results indicate that plasticity (in the presence of elasticity) can enhance momentum transport and that the nonlinear interplay between plasticity and elasticity can lead to counterintuitive flow behavior (e.g., non-monotonic dependence of $\phi$ on $Bn$). More broadly, our simulations indicate that materials that possess both yield-stress and elasticity such as muds \& soils, clay suspensions, and gels are prone to hydrodynamic instabilities even when yield-stresses are large relative to elastic stresses. 
\begin{acknowledgments}
We appreciate helpful conversations with Rob Poole, Alex Morozov, Monica Naccache, and Paulo de Souza Mendes. 
\end{acknowledgments}

\bibliography{refs.bib}

\end{document}